# Hints vs Distractions in Intelligent Tutoring Systems: In search of the proper type of help

**Maria Blancas-Muñoz**
**Vasiliki Vouloutsi**
**Riccardo Zucca**
**Anna Mura**
Synthetic Perceptive Emotive Cognitive Systems (SPECS) group, Institute for Bioengineering of Catalonia (IBEC)

mblancas@ibecbarcelona.eu
vvouloutsi@ibecbarcelona.eu
rzucca@ibecbarcelona.eu
amura@ibecbarcelona.eu

**Paul F.M.J. Verschure**
Synthetic Perceptive Emotive Cognitive Systems (SPECS) group, Institute for Bioengineering of Catalonia (IBEC), Catalan Institution for Research and Advanced Studies (ICREA)

pverschure@ibecbarcelona.eu

## Abstract
The kind of help a student receives during a task has been shown to play a significant role in their learning process. We designed an interaction scenario with a robotic tutor, in real– life settings based on an inquiry-based learning task. We aim to explore how learners' performance is affected by the various strategies of a robotic tutor. We explored two kinds of (presumable) help: hints (which were specific to the level or general to the task) or distractions (information not relevant to the task: either a joke or a curious fact). Our results suggest providing hints to the learner and distracting them with curious facts as more effective than distracting them with humour.

## ACM Classification Keywords
I.2.11 Intelligent agents, K.3.1 Computer-assisted instruction

## Introduction
Intelligent Tutoring Systems (ITS) provide an opportunity to monitor students' learning process and generate consequent learning experiences adapted to their differences and needs. To do so, a pedagogical model of the learner is needed to control the ITS. We propose the Distributed Adaptive Control (DAC) pedagogical model, which establishes the phases of learning the student goes through before obtaining new knowledge and how can they be used to scaffold their learning process, as explained in [1].

When developing an ITS, it is important to pay attention to the content provided (for example the difficulty of the task). However, equally important is knowing what the system should do when the learner finds themselves in a situation where they need help. This study aims to explore how performance is affected by the various strategies of a robotic tutor to provide help to the learner in dyadic

> **The Balance-Beam Scenario**
>
> In the balance beam scenario, different numbers of weights are placed at varying distances from the fulcrum on equally spaced pegs positioned on both arms of the scale. The puzzles we provided include two weights (red and yellow, the latter being twice as heavy as the red) and have four levels of increased difficulty, matching Siegler's rules:
>
> - **Level I:** different weights are placed at the same distance from the centre of the balance
> - **Level II:** equal weights are placed at different distances from the centre of the balance
> - **Level III:** different weights are placed at different distances from the centre of the balance
> - **Level IV:** following the same principles of Level III, the number of weights at each side varies interactions. We decompose the help strategies into two main categories: hints and distractions as we want to assess the differences between helping the learner by providing them with information about the task and by making them stop what they are doing to do something different.

Providing help adapted to the student's needs can promote learning and higher performance as well as a more accurate self-monitoring of their abilities and skills [2]. A common help strategy is an explanation of how to execute correctly a task or the provision of information that is relevant to the specific content. Although various help mechanisms have been explored in learning environments, the role of distraction has not been sufficiently examined.

Distraction is the process of shifting one's focus to events or stimuli that block or diminish the acquisition of desired information. Distractions can be internal or external and can be relevant or irrelevant to a specific task. It seems that congruent (or relevant) distraction facilitates performance, increases response times, reduces forgetting [3] and opposite effects are observed when distraction is incongruent. Similarly, studies on interleaving (where material from different categories is presented mixed) have resulted in enhanced inductive learning compared to temporal spacing, where the differences between categories are not highlighted [4].

We designed an interaction scenario in real–life settings based on an inquiry-based learning task. Typically, inquiry-based learning tasks involve active exploration of the world by asking questions, making discoveries and testing hypotheses. The proposed scenario aimed at teaching children about physics concepts based on the Piagetian balance-beam experiments [5] matching Siegler's rules [6] (explained in the box at the left).

Children explore the physics of the balance problem guided by an artificial agent (robot) and interact with a virtual balance via an Android tablet. The goal of the interaction is that the child learns about balance and momentum by going through a series of puzzle tasks with the balance beam. The artificial agent is there to encourage the students, to help them get through the different tasks and to provide feedback; thus, learning is supported by continuously monitoring the learner's progress and generating exercises and hints tailored to their needs.

Robots' social abilities and skills make them relevant for peer-to-peer interaction [7] as they may influence children's knowledge acquisition. For example, the presence of a robot (compared to a screen) may account for higher learning gains [8], [9], whereas the role assumed by the robot (peer or tutor) has been examined in various educational contexts [10], [11].

### Methodology
Here we report a six weeks-long in-school study that evaluates the effects of help provided to the learner by an artificial tutor while performing a scientific inquiry-learning task. In total, 60 children (29 female) participated, aged between 8 – 9 years old. The participants performed the task individually, and the design was between subjects, so each participant only received one kind of help. The task was designed to adapt to the required difficulty of the learner by increasing or decreasing the level of the provided exercises depending on their performance.

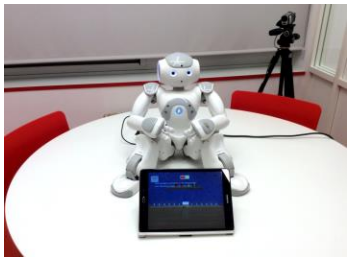

Figure 1: The experimental setup. The tablet presents the child with the different exercises and a Nao robot assists the interaction. Every session was recorded with a video camera, actions from the robot and inputs provided by the children through the device were logged for online and offline analysis

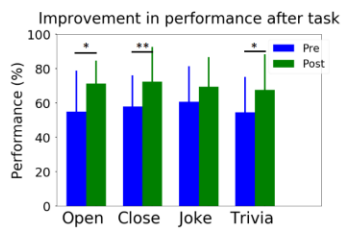

Figure 2: Improvement in performance after the task. When comparing the differences between pre and post tests for each condition, we can see that performance improves in all the conditions but the "Joke" one. No differences were found among conditions in performance in pre-test.

The robot provided two kinds of help: hints or distractions. Hints were further subdivided to "Open" (16 participants) and "Closed" (14 participants). Open hints were relevant to the task in general (e.g. "Remember that the yellow weight is two times heavier than the red weight"). Closed hints were specific to the difficulty of each level (e.g. "Remember that if the distance is the same, what matters is the weight" for level I, "Remember that if the weights are the same, what matters is the distance" for level II etc.). We divided distractions into two subcategories: "Trivia" (13 participants) and "Jokes" (17 participants). Trivia refers to knowledge-related facts such as "Did you know that the male seahorse is the one that gets pregnant?". Jokes are funny stories like "What did the traffic light say to the other? - Do not look at me; I am changing!". The provided jokes were appropriate for the age group of the children. Participants in the pilot tests provided several of the jokes told by the robot. On average, we made sure that the robot's utterances lasted approximately the same amount of time to avoid any biases.

The experimental sessions consisted of five main phases. Before the main experiment, children were first introduced to the purpose of the task and filled in a pre-assessment questionnaire. Then children interacted with the Nao robot while performing the balance beam task using a virtual reality application with a tablet (see Figure 1). Finally, after the main experiment, children had to fill in a post-assessment questionnaire and a semi-structured interview to evaluate the interaction and the task. All participants provided with a consent form from their parents, and the study was approved by a local ethical committee.

The robot (in all conditions) positively encouraged the learner in correct and incorrect answers by telling them "Well done!" or "Do not worry, you will do better next time". The control architecture of the proposed setup is described in detail in [1]. For each exercise, children had to report their confidence level before viewing the answer on the device. The maximum number of exercises was capped at 24. However, from the 16th trial, the synthetic tutor would ask the student if they wanted to continue and the child provided the answer via the tablet.

**Results**

When exploring the results of comparing the performance of the pre-test with the one of the post-test, we observe a significant improvement in performance for the participants in the "Open" (pre: 60.71±21.92, post:72.62±12.37), "Close" (pre: 58.33±21.25, post:72.22±21.61) and "Trivia" (pre: 55.88±21.34, post:67.65±20.19) conditions (Figure 2). Both conditions related to providing help to the participant on the content of the task ("Open", "Close") and the "Trivia" one show strong significant differences between pre and post ($t(15)=-2.19, p=0.047$; $t(13)=-3.16, p=0.009$; $t(16)=-2.426, p=0.027$; respectively). No differences were found among conditions in performance in the pre-test.

Being an adaptive-difficulty task, we analysed on-task performance as the percentage of trials in level 1 (as the difficulty has not been adapted yet in these trials). A Kruskal-Wallis H-test showed significant differences between conditions in the percentage of level 1 trials ($h = 8.69, p = 0.03$). More specifically, the "Trivia" (Median: 16.67, Median Absolute Deviation -MAD-: 4.17) condition was the one with a lower percentage of level 1 trials, being significantly different from the "Joke" (Median: 41.67, MAD: 22.96) and "Open" (Median: 29.17, MAD: 10.42) conditions (Mann-

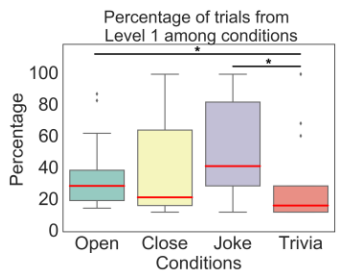

Figure 3: Number of level 1 trials per condition. When comparing the differences in number of trials of the first level for each group, we can see that the "Trivia" condition is the one with a lower number, being significantly different from the "Open" and "Joke" conditions.

## Acknowledgements
This work is supported by grants from the European Research Council under the European Union's 7th Framework Programme FP7/2007-2013/ERC grant n.341196 (CDAC) and the European Commission's Horizon 2020 socSMC grant (no. socSMC-641321H2020-FETPROACT-2014) to Paul F.M.J.Verschure. We would like to thank the participants, their parents and teachers for their contribution.

Whitney, $h=50$, $p=0.006$ and $h=74$, $p=0.01$, respectively), as shown in Figure 3.

## Conclusion and Discussion

The present study examines the effect of several kinds of help on learners' performance and the differences between them. Four types of help were provided: two types of hints about the task's content and two types of distractions from the task process. Participants carried out a science-based task where a robot helped them with one of the four types of help.

As it could be expected, our results suggest that providing learners with hints about the task (compared to distracting them with jokes) seem to be more effective to increase their performance. Nevertheless, they seem not to be the only effective type of help, as distracting learners with curious facts also relates to an increase in performance. The results about performance during the task seem to be in line with this last finding, as the "Trivia" condition was the one with less amount of level-one trials, suggesting that participants in this condition advanced faster to more difficult exercises compared to the other conditions.

These results suggest distraction as a possible type of help to consider in an ITS. Further research will analyse how its effect varies depending on individual differences of the learner, for example: their gender, metacognitive abilities or emotional state during the task.